\documentclass[11pt]{article}

\usepackage[preprint]{acl}
\usepackage{etoolbox}
\usepackage{multirow}
\usepackage{hyperref}
\usepackage{url}
\usepackage{multirow}
\usepackage[table]{xcolor}
\usepackage{color}
\usepackage{booktabs}
\usepackage{subcaption} 
\usepackage{changepage} 
\usepackage{listings}
\usepackage{caption}
\usepackage{enumitem}
\usepackage{xspace}
\usepackage{tcolorbox}
\usepackage{enumitem}
\usepackage{xcolor}
\usepackage{listings}
\usepackage{tcolorbox}
\tcbuselibrary{most}
\usepackage{times}
\usepackage{latexsym}
\usepackage{balance}
\usepackage{bm}
\usepackage[table]{xcolor} 
\usepackage{colortbl} 
\usepackage{amsmath, amsthm, amssymb, amsfonts} 
\usepackage{tikz} 
\usetikzlibrary{calc} 

\DeclareCaptionFont{white}{\color{white}}
\DeclareCaptionFormat{listing}{%
  \colorbox{gray}{\parbox{\dimexpr\linewidth-2\fboxsep\relax}{#1#2#3}}%
}

\usepackage[absolute,overlay]{textpos}

\definecolor{lightgray}{RGB}{245,245,245}  
\definecolor{darkgray}{RGB}{80,80,80}      
\definecolor{bordergray}{RGB}{180,180,180} 
\definecolor{textcolor}{RGB}{50,50,50}     
\definecolor{linegray}{RGB}{200,200,200}     

\lstdefinestyle{largestyle}{
    basicstyle=\ttfamily\footnotesize\color{textcolor},
    breaklines=true,
    breakindent=0pt,
    frame=none,
    backgroundcolor=\color{lightgray},
    tabsize=2,
    showstringspaces=false,
    columns=flexible,
    keepspaces=true,
    aboveskip=5pt,      
    belowskip=5pt       
}

\newtcolorbox{simplebox}[2][]{
    enhanced,
    breakable,
    colback=lightgray,
    colframe=bordergray,
    frame style={line width=0.5pt},
    colbacktitle=darkgray,
    coltitle=white,
    fonttitle=\sffamily\bfseries\scriptsize,,
    attach boxed title to top left={xshift=10pt,yshift=-2.5mm},
    boxed title style={
        colframe=darkgray,
        colback=darkgray,
        boxrule=0.5pt,
        arc=0pt,
        top=0.5pt,    
        bottom=0.5pt
    },
    left=12pt,
    right=12pt,
    top=0pt,            
    bottom=0pt,       
    before upper={\vspace{1.5mm}},
    after upper={\vspace{-1.5mm}},
    title={#2},
    #1
}
\usepackage[T1]{fontenc}
\usepackage[utf8]{inputenc}

\usepackage{microtype}

\usepackage{inconsolata}

\usepackage{graphicx}

\newcommand{\name}{OneRec-Think\xspace}

\title{\textsc{OneRec-Think}: In-Text Reasoning for Generative Recommendation}

\makeatletter
\def\thanks#1{\protected@xdef\@thanks{\@thanks
        \protect\footnotetext{#1}}}
\makeatother

\author{Zhanyu Liu\thanks{*: Equal contributions.}\textsuperscript{*}, 
Shiyao Wang\textsuperscript{*}, 
Xingmei Wang,
Rongzhou Zhang,
Jiaxin Deng,
\\\textbf{
Honghui Bao,
Jinghao Zhang,
Wuchao Li,
Pengfei Zheng,
Xiangyu Wu,
}
\\\textbf{
Yifei Hu,
Qigen Hu,
Xinchen Luo,
Lejian Ren,
Zixing Zhang,
}
\\\textbf{
Qianqian Wang,
Kuo Cai,
Yunfan Wu,
Hongtao Cheng,
Zexuan Cheng,
}
\\\textbf{
Lu Ren,
Huanjie Wang,
Yi Su,
Ruiming Tang,
Kun Gai,
Guorui Zhou\thanks{\textdagger{}: Corresponding Author.}\textsuperscript{\textdagger}}
\\
Kuaishou Inc., Beijing, China\\
\texttt{\{liuzhanyu,wangshiyao08,zhouguorui\}@kuaishou.com}\\
}

\begin{document}

\maketitle

\begin{tikzpicture}[remember picture,overlay]
    \node[anchor=north west,inner sep=0] at ($(current page.north west)+(1.5cm,-1.2cm)$) 
        {\includegraphics[height=0.8cm]{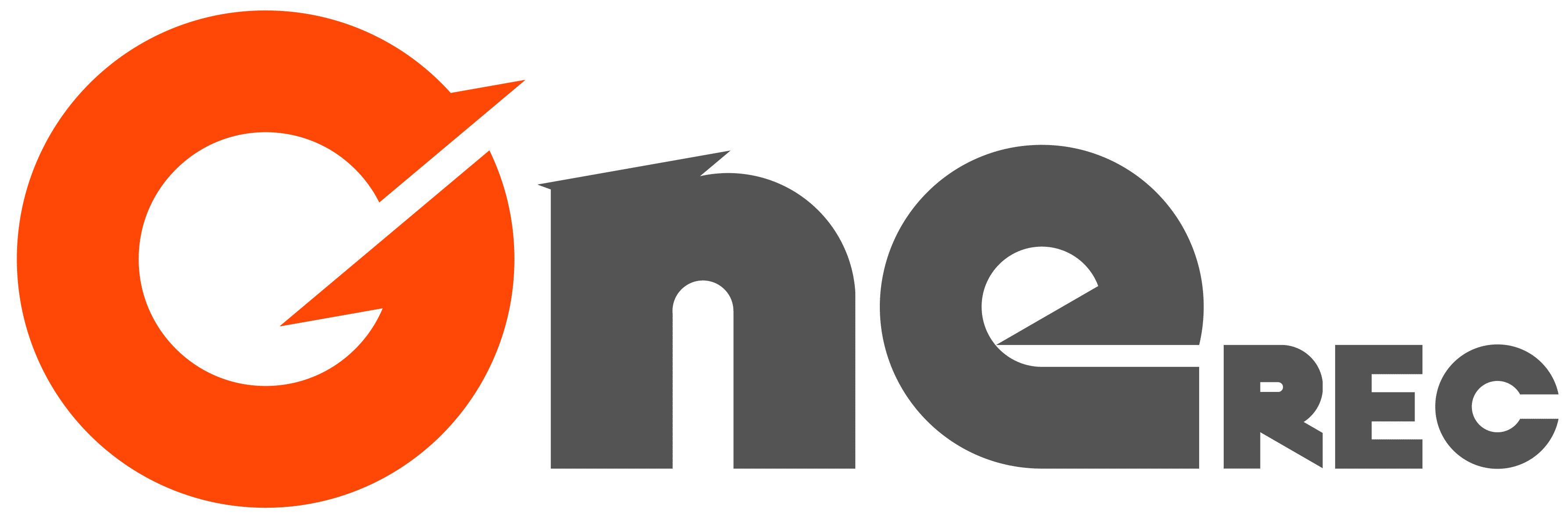}};
    
    \coordinate (logo base) at ($(current page.north west)+(1.5cm,-1.2cm-0.9cm)$);
    
    \def\LeftMargin{0.5cm}   
    \def\RightMargin{1cm} 
    
    \draw[linegray,line width=0.8pt] 
        ($(logo base) - (\LeftMargin,0)$)  
        -- 
        ($(logo base -| current page.east) - (\RightMargin,0)$); 
\end{tikzpicture}
\begin{abstract}
The powerful generative capacity of Large Language Models (LLMs) has instigated a paradigm shift in recommendation. However, existing generative models (e.g., OneRec) operate as implicit predictors, critically lacking the capacity for explicit and controllable reasoning—a key advantage of LLMs. To bridge this gap, we propose \name, a unified framework that seamlessly integrates dialogue, reasoning, and personalized recommendation. \name incorporates: (1) \textbf{Itemic Alignment}: cross-modal Item-Textual Alignment for semantic grounding; (2) \textbf{Reasoning Activation}: Reasoning Scaffolding to activate LLM reasoning within the recommendation context; and (3) \textbf{Reasoning Enhancement}, where we design a recommendation-specific reward function that accounts for the multi-validity nature of user preferences. Experiments across public benchmarks show state-of-the-art performance. Moreover, our proposed "Think-Ahead" architecture enables effective industrial deployment on Kuaishou, achieving a 0.159\% gain in APP Stay Time and validating the practical efficacy of the model's explicit reasoning capability.
\end{abstract}

\section{Introduction}

\begin{figure}[!th]
\centering
\includegraphics[width=0.9\columnwidth]{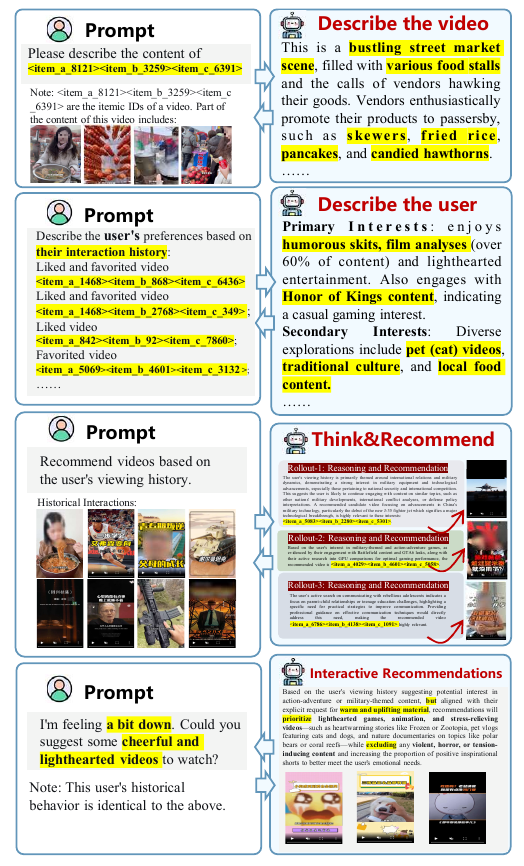} 
\caption{Examples of \name's Unified Dialogue, Reasoning and Recommendation Framework.}
\vspace{-.5cm}
\label{fig:demo}
\end{figure}

The rapid advancement of Large Language Models (LLMs) has fundamentally reshaped recommender systems, ushering in the generative retrieval paradigm(GR)~\citep{peng2025survey,zhang2025survey,deldjoo2024review,li2023large,wang2023generative}. This approach represents a profound shift from traditional query-candidate matching, utilizing Transformer-based sequence-to-sequence models to autoregressively decode the identifiers of target candidates. Capitalizing on this, a major research frontier is the development of end-to-end generative frameworks, including OneRec, OneLoc, OneSug, and OneSearch~\citep{deng2025onerec,zhou2025onerec1,zhou2025onerec2,wei2025oneloc,guo2025onesug,chen2025onesearch}. These unified models replace the traditional multi-stage recommendation funnel (involving separate retrieval and ranking stages), enabling holistic optimization towards the final objective and concentrating computational resources for better industrial scaling and performance.

While these models successfully harness the LLMs' capacity for output generation, they fundamentally lack the explicit, verifiable reasoning pathways that define modern LLM breakthroughs, such as text-based Chain-of-Thought (CoT)~\citep{rajput2023recommender,zheng2023adapting,wang2024content}.
To bridge this critical gap, we propose \textbf{\name, a novel framework that integrates dialogue, reasoning, and personalized generative recommendations within a single, unified model}. It is capable of generating high-quality, interpretable textual reasoning paths, significantly enhancing both recommendation accuracy and user trustworthiness. The model's inherent dialogic nature further enables dynamic tailoring of suggestions to specific user constraints (as shown in Fig.~\ref{fig:demo}).
Our approach is realized through a three-stage framework: 
(1) \textbf{Itemic Alignment}, which maps item semantics into the LLM's textual embedding space, establishing a unified representational continuum that unlocks the model's capacity for reasoning.
(2) \textbf{Reasoning Activation}, which aims to induce the LLM's inherent reasoning ability directly within the context of recommender systems;
(3) \textbf{Reasoning Enhancement}, which utilizes a recommendation-specific reward function that  captures the multi-validity (i.e., multiple valid choices) nature of user preferences. Furthermore, we introduce the \name Inference Architecture to ensure efficient deployment and real-time responsiveness in industrial-scale serving scenarios.
Our contributions are summarized as follows: 
\begin{itemize}
\setlength{\itemsep}{0.5ex}  
\setlength{\parskip}{0pt}    
\item[\textbullet] We introduce a unified framework that bridges the semantic gap between discrete recommendation items and continuous reasoning spaces, enabling seamless integration of personalized recommendation within LLMs' natural language understanding.
\item[\textbullet] We design a novel reasoning paradigm that orchestrates multi-step deliberation with recommendation optimization, achieving interpretable and accuracy-aware personalized recommendation through synergistic training.
\item[\textbullet] The proposed approach achieves state-of-the-art results on multiple public benchmarks, while our deployment-friendly "Think-Ahead" architecture enables significant industrial impact with a 0.159\% gain in APP Stay Time.
\end{itemize}

\section{Related Work}
\subsection{Reasoning in Large Language Models}
Large language models achieve complex reasoning through various prompting techniques, with CoT prompting~\citep{wei2022chain} being the foundational approach that decomposes problems into intermediate reasoning steps. 
This has inspired numerous extensions including zero-shot CoT~\citep{kojima2022large}, self-consistency decoding~\citep{wang2022self}, and tree-of-thoughts~\citep{yao2023tree}. 
These techniques enable test-time scaling where additional computational budget during inference improves performance~\citep{snell2024scaling}.
Recent work has shifted focus from prompting to post-training enhancement of reasoning capabilities by using techniques such as Reinforcement Learning.
Models including DeepSeek-R1~\citep{guo2025deepseek} and Seed-1.5~\citep{seed2025seed1} optimize reasoning behaviors via techniques such as GRPO~\citep{shao2024deepseekmath}, DAPO~\citep {yu2025dapo}, and VAPO~\citep {yue2025vapo}, which demonstrates promising advancements in this direction.

\subsection{Reasoning-Based Recommendation}
Although generative recommendation models such as TIGER~\citep{rajput2023recommender}, HSTU~\citep{zhai2024actions}, and OneRec~\citep{zhou2025onerec1} have demonstrated effectiveness, they inherently lack reasoning capabilities. 
Recently, the Reasoning-based recommendation systems aim to perform multi-step deduction for more accurate and interpretable recommendations.
Existing approaches fall into two categories: \textbf{explicit reasoning} methods generate human-readable rationales but are confined to discriminative tasks~\citep{RecSAVER,reasoningrec,DeliberativeRec,exp3rt}, while \textbf{implicit reasoning} methods~\citep{slowthink,tang2025think} perform latent reasoning without textual interpretability.
Our work introduces explicit reasoning into generative recommendation, bridging this gap to enable both interpretable rationales and scalable item generation.

\begin{figure*}[!tb]
    \centering 
    \includegraphics[width=1.0\textwidth]{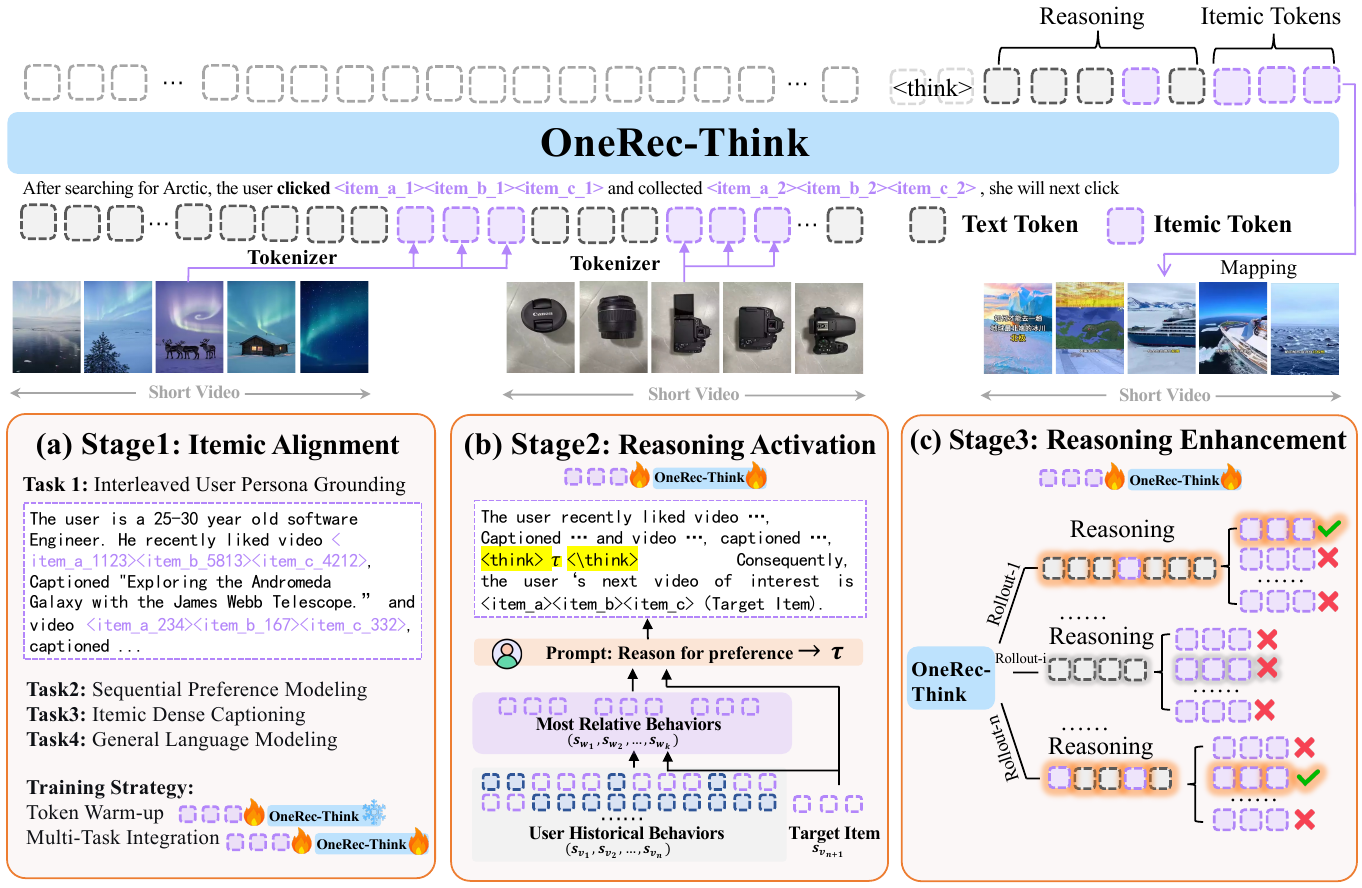} 
    \caption{The framework of the \name. In the first stage, we achieve item-level semantic alignment through multi-task pre-training. In the second stage, we activate explicit reasoning by prompting the model to generate preference rationales. In the third stage, we refine the reasoning paths through RL based on a reward tailored for recommendations.}
    \label{fig:main} 
\end{figure*}

\section{Preliminary}
\paragraph{Itemic Token}
An itemic token is a discrete, semantic-rich representation unit for an item, analogous to a word token in natural language. 
Following OneRec~\citep{zhou2025onerec1,rajput2023recommender}, we map each item $v$ to a sequence of such tokens $\bm{s}_v = (s_v^1, \dots, s_v^L)$, which are generated from the item's multi-modal and collaborative content. 

\paragraph{Problem Definition}
Let $\mathcal{U}$ and $\mathcal{V}$ denote the sets of users and items, respectively. Each user $u \in \mathcal{U}$ has a chronological interaction history $V_u = (v^u_1, v^u_2, \dots, v^u_{n_u})$ of length $n_u$. 
For brevity, we omit the user index $u$. 
By using the itemic tokens, the user's interaction history is thus represented by the sequence $S_u=(\bm{s}_{v_1}, \dots, \bm{s}_{v_{n}})$.

Conventional generative recommenders~\citep{rajput2023recommender, zhou2025onerec1, zhou2025onerec2, zheng2023adapting} define their task with a generation target of the next itemic tokens as:
\begin{equation}
    \bm{s}_{v_{n+1}} \sim P(\cdot|\bm{s}_{v_1}, \dots, \bm{s}_{v_{n}};\theta)
\end{equation}
In contrast, we reformulate this task to unify reasoning and recommendation in a single autoregressive pass. Conditioned on a prompted user history, we generate tokens sequentially, beginning with a reasoning sequence $\bm{\tau}=(r_1, \dots, r_M)$ and concluding with the next itemic tokens $\bm{s}_{v_{n+1}}$. This end-to-end process is captured by:
\begin{equation}
\begin{aligned}
    \bm{\tau} &\sim P\left(\cdot \mid \mathcal{P}(\bm{s}_{v_1}, \dots, \bm{s}_{v_n}); \theta\right) \\
\bm{s}_{v_{n+1}} &\sim P\left(\cdot \mid \mathcal{P}(\bm{s}_{v_1}, \dots, \bm{s}_{v_n}), \bm{\tau}; \theta\right)
\end{aligned}
\end{equation}
where $\mathcal{P}(\cdot)$ means a valid prompt constructed for the recommendation.

\section{Methodolody}

We now present \name, a scalable framework for end-to-end generative reasoning recommendation. Our approach comprises three core components: an Itemic Alignment stage, a Reasoning Activation stage, and a "Think Ahead" architecture for industrial deployment.
The illustration of \name is shown in Fig.~\ref{fig:main}.

\subsection{Itemic Alignment through Multi-Task Pre-training}\label{sec:semantic_alignment}
To align recommender knowledge with the LLM's linguistic space, we design a Multi-Task Pre-training strategy enabling seamless processing of natural language and itemic tokens via four complementary tasks under Next Token Prediction.

\paragraph{Interleaved User Persona Grounding}
Unlike prior work that uses either pure textual data or isolated item sequences, this task interleaves the itemic tokens and text tokens of User Persona.
It includes serialized static attributes, active search behaviors, interactive sequences, and summarized user interests. 
This composition creates rich, dual-modality training instances where itemic tokens are grounded in their semantic context.

\paragraph{Sequential Preference Modeling}
As the core recommendation task, this task constructs data that teaches the model to predict subsequent item interactions from chronological user histories.

\paragraph{Itemic Dense Captioning}
This task requires the model to decode an item's descriptive content from its itemic tokens.
By learning to generate detailed textual descriptions, the model establishes a fundamental understanding of the semantic characteristics represented by item combinations.

\paragraph{General Language Modeling}
This task continues pre-training the model on general text corpora, preserving the model's fundamental language capabilities during applying the model to recommendation scenarios.

To enable effective knowledge integration while preserving the linguistic capabilities of the model, we implement a two-substage training strategy to ensure stable alignment. 
The \textbf{Token Warm-up} substage exclusively trains itemic token embeddings on the Interleaved User Persona Grounding task while keeping the base LLM frozen.
The subsequent \textbf{Multi-Task Integration} substage jointly optimizes all parameters on the combined task using a designed ratio (see Appendix~\ref{app:industry_setting} for details).

\subsection{Reasoning Activation}

Despite robust itemic alignment, direct application to industrial recommendation scenarios often fails to yield effective CoT reasoning due to the noisy and lengthy nature of real-world user behavior sequences.
To address this, we propose a supervised fine-tuning framework that first extracts coherent reasoning trajectories from pruned user contexts, then leverages these trajectories to guide rationale generation over raw behavioral data, enabling effective contextual distillation for noisy industrial settings (as shown in Fig.~\ref{fig:main}(b)).

\noindent{\textit{\textbf{Bootstrapping with Pruned Contexts:}}}
To bootstrap reasoning capabilities, we first construct easy-to-learn instances where logical relationships are preserved despite sequence pruning. For each user, we select the target item $\bm{s}_{v_{n+1}}$ and form a context-target pair $<(\bm{s}_{v_1},\dots,\bm{s}_{v_{n}}), \bm{s}_{v_{n+1}}>$. We then retrieve the top-$k$ most relevant historical items using a similarity function $g(\cdot,\cdot)$:
\begin{equation}
g((\bm{s}_{v_1},\dots,\bm{s}_{v_{n}}), \bm{s}_{v_{n+1}}) = (\bm{s}_{w_1},\dots,\bm{s}_{w_k}).
\end{equation}
Using these relevant items, we query our pre-aligned model to generate a rationale $\bm{\tau}$ explaining the target interaction:
\begin{equation}
\begin{aligned}
\bm{\tau} &\sim P\left(\cdot \mid \mathcal{P}_{r}((\bm{s}_{w_1},\dots,\bm{s}_{w_k}), \bm{s}_{v_{n+1}}); \theta\right)
\end{aligned}
\end{equation}
where $\mathcal{P}_{r}(a,b)$ means constructs a prompt to query the rationale why a user who interacts with item sequence $a$ would interact with item $b$.
This process yields high-quality rationales that are both logically sound and target-aligned, providing ideal training signals for reasoning induction.

\noindent{\textit{\textbf{Learning to Reason from Noisy Sequences:}}}
The distilled rationales serve as supervision for learning to reason from raw sequences. 
The training objective minimizes the negative log-likelihood of generating both the rationale and target item:
\begin{equation}
\label{eq:sft_loss}
\begin{aligned}
\mathcal{L}_{\text{RA}} = - \Biggl( 
     \sum_{i=1}^{M} \log P(r_i|\mathcal{P}(\bm{s}_{v_1}, \dots, \bm{s}_{v_n}), r_{<i};\theta) \\
     + \sum_{j=1}^{L} \log P(s_{v_{n+1}}^j | \mathcal{P}(\bm{s}_{v_1}, \dots, \bm{s}_{v_{n}}), \bm{\tau}, {s}_{v_{n+1}}^{<j};\theta) \Biggr),
\end{aligned}
\end{equation}
where $\bm{\tau} = \{r_1,\dots,r_M\}$ represents the rationale tokens and $\bm{s}_{v_{n+1}} = \{s_{v_{n+1}}^1,\dots,s_{v_{n+1}}^L\}$ denotes the target item tokens.
By optimizing $\mathcal{L}_{\text{RA}}$, the model learns to internally distill relevant context from noisy sequences and generate coherent rationales that bridge user history to target interactions, significantly enhancing its CoT capabilities in challenging recommendation scenarios.

\subsection{Reasoning Enhancement}

Building upon the CoT capabilities from Reasoning Activation, we address the challenge of ensuring consistently high-quality reasoning through Reinforcement Learning. 
This stage refines the recommendation accuracy using a novel reward mechanism tailored for generative recommendation.

\noindent{\textit{\textbf{Beam Candidate Reward Maximization:}}}
Standard verifiable pass rewards face significant sparsity challenges in recommendation scenarios, as most reasoning rollouts fail to hit the target item and consequently yield identical zero rewards—thereby neutralizing group advantages in algorithms such as GRPO~\citep{shao2024deepseekmath}.
To overcome this, we introduce the \textbf{Rollout-Beam} reward that evaluates reasoning capability by the model's best achievable performance within a constrained beam.
Our approach employs beam search with width $K$ to explore multiple generation candidates after reasoning trajectory generation:
\begin{equation}
\label{eq:reward}
    \mathcal{R}_{\text{Rollout-Beam}} = \max_{\hat{s}_{v_{n+1}} \in \mathcal{B} }\sum_{l=1}^{L} \mathbb{I}(\hat{s}_{v_{n+1}}^l = s_{v_{n+1}}^l),
\end{equation}
where the beam search result set is defined as :
\begin{equation}
\begin{split}
\mathcal{B} &= \bigl\{ \bigl( \hat{s}_{v_{n+1}}^{1,(j)},\cdots, \hat{s}_{v_{n+1}}^{L,(j)} \bigr) _{j=1}^K\bigr\}, \\
            &= \text{BeamSearch}\Bigl(P(\bm{s}_{v_{n+1}} \mid \bm{H}, \bm{\tau};\theta\bigr), K\Bigr)
\end{split}
\end{equation}
which contains the items with the top-$K$ probabilty in the beam search within the distribution $P\left(\bm{s}_{v_{n+1}} \mid \bm{H}, \bm{\tau};\theta\right)$.
$\bm{H} = \mathcal{P}(\bm{s}_{v_1}, \dots, \bm{s}_{v_n})$ is a valid prompt of history sequence.
$\text{BeamSearch} (P,K)$ means the top-$K$ result of beam search within distribution $P$.
Subsequently, we optimize the model using GRPO~\citep{shao2024deepseekmath} based on $ \mathcal{R}_{\text{Rollout-Beam}}$, which effectively leverages the enriched reward signals from the multi-validity nature of user preferences. 

Overall, this design establishes training-inference consistency by aligning reward computation with beam search-based inference, providing denser learning signals through multi-path evaluation.

\begin{table*}[!th]
\setlength{\abovecaptionskip}{0.05cm}
\setlength{\belowcaptionskip}{0.2cm}
\caption{Overall performance comparison between the baselines and \name on three datasets. The bold results highlight the best results, while the second-best ones are underlined.}

\setlength{\tabcolsep}{2mm}{
\resizebox{\textwidth}{!}{
\setlength{\extrarowheight}{0pt}
\resizebox{0.7\linewidth}{!}{
\begin{tabular}{c|cccccccc>{\columncolor{blue!6}}c}
\toprule
\textbf{Dataset}  & \textbf{Method} & \textbf{BERT4Rec} & \textbf{HGN} & \textbf{GRU4Rec} & \textbf{SASRec} & \textbf{TIGER}  & \textbf{HSTU} & \textbf{ReaRec}& \textbf{\name} \\
\midrule
\multirow{4}{*}{\textbf{Beauty}} & R@5 & 0.0232 & 0.0319 & 0.0395 & 0.0402 & 0.0405  & 0.0424 & \underline{0.0450} & \textbf{0.0563} \\
 & R@10 & 0.0396 & 0.0536 & 0.0584 & 0.0607 & 0.0623 &  0.0652 & \underline{0.0704} & \textbf{0.0791} \\
 & N@5 & 0.0146 & 0.0196 & 0.0265 & 0.0254 & 0.0267  & \underline{0.0280} & 0.0262 & \textbf{0.0398} \\
 & N@10 & 0.0199 & 0.0266 & 0.0326 & 0.0320 & 0.0337 & \underline{0.0353} & 0.0344 & \textbf{0.0471} \\
\midrule
\multirow{4}{*}{\textbf{Sports}} & R@5 & 0.0102 & 0.0183 & 0.0190 & 0.0199 & 0.0215  & \underline{0.0268} & 0.0214 & \textbf{0.0288} \\
 & R@10 & 0.0175 & 0.0313 & 0.0312 & 0.0301 & \underline{0.0347}  & 0.0343 & 0.0332 & \textbf{0.0412} \\
 & N@5 & 0.0065 & 0.0109 & 0.0122 & 0.0106 & 0.0137  & \underline{0.0173} & 0.0116 & \textbf{0.0199} \\
 & N@10 & 0.0088 & 0.0150 & 0.0161 & 0.0141 & 0.0179 &  \underline{0.0226} & 0.0154 & \textbf{0.0239} \\
\midrule
\multirow{4}{*}{\textbf{Toys}} & R@5 & 0.0215 & 0.0326 & 0.0330 & 0.0448 & 0.0337  & 0.0366 & \underline{0.0523} & \textbf{0.0579} \\
 & R@10 & 0.0332 & 0.0517 & 0.0490 & 0.0626 & 0.0547  & 0.0566 & \underline{0.0764} & \textbf{0.0797} \\
 & N@5 & 0.0131 & 0.0192 & 0.0228 & \underline{0.0300} & 0.0209  & 0.0245 & 0.0298 & \textbf{0.0412} \\
 & N@10 & 0.0168 & 0.0254 & 0.0279 & 0.0358 & 0.0276  & 0.0309 & \underline{0.0376} & \textbf{0.0482} \\
\bottomrule
\end{tabular}
}}}
\label{tab:MainTable}
\end{table*}

\subsection{Industrial Deployment: A "Think-Ahead" Architecture}

The deployment of \name in industrial recommendation systems presents a fundamental challenge: reconciling the computational demands of multi-step reasoning with the stringent latency requirements of real-time user interactions. 

To address this critical bottleneck, we introduce a novel \emph{Think-Ahead} Inference Architecture. Our solution strategically decouples the model's inference into two stages: \textbf{In the first stage}, the computationally intensive reasoning path and the initial item-tokens (e.g., the first two itemic tokens) are generated offline by the full \name model. These initial tokens are designed to capture the user's broad intent or general preference context. Subsequently, \textbf{the second stage} then employs a real-time updated OneRec model following \cite{zhou2025onerec1} for online finalization. It utilizes the pre-generated item-tokens as a constrained prefix to rapidly produce the final itemic token. This design ensures real-time responsiveness and achieves production-grade performance by leveraging current contextual data.
The details of this architecture are in Appendix~\ref{app:think_ahead}.

\section{Experiments}\label{sec:exp}

\begin{figure*}[!t]
    \centering
    \includegraphics[width=1\textwidth]{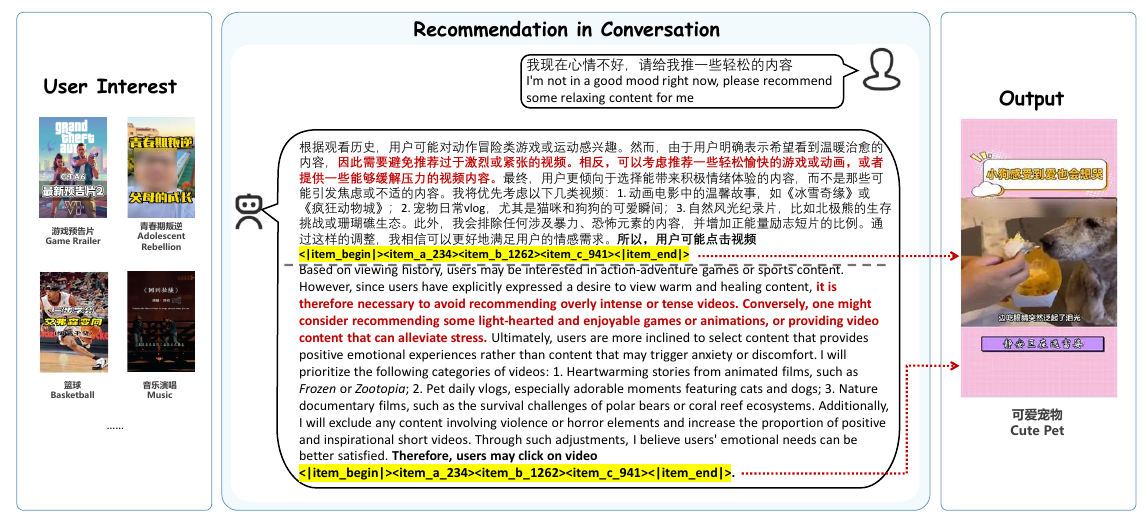}
    \caption{Demonstration of context-aware recommendation adaptation: our model dynamically shifts recommendations to relaxing content based on the user's command.}
    \label{fig:case_CoT_3}
\end{figure*}

\begin{figure*}[!t]
    \centering
    \includegraphics[width=1\textwidth]{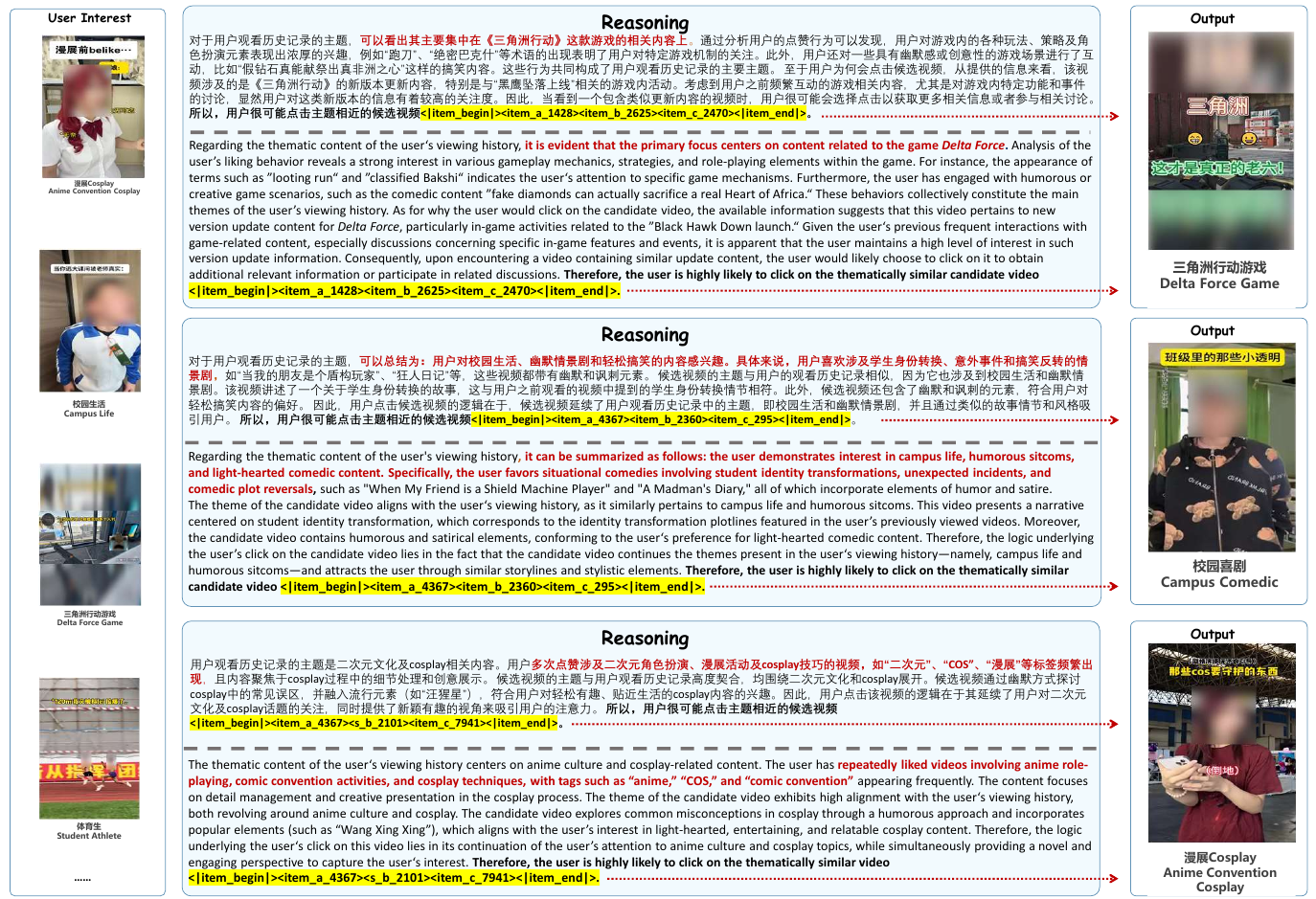}
    \caption{Demonstration of fine-grained interest reasoning, which shows the end-to-end process from user behavior analysis to interpretable recommendations.}
    \label{fig:case_CoT_4}
\end{figure*}

\subsection{Experimental Settings}\label{sec:exp_setting}

\textbf{Datasets and Baselines.}

We use three real-world recommendation datasets from the popular Amazon review benchmark\footnote{\url{https://jmcauley.ucsd.edu/data/amazon/}.}: Beauty, Toys, and Sports.
We compare \name against two groups of competitive baselines: (1) Classic sequential methods like BERT4Rec~\citep{sun2019bert4rec}, HGN~\citep{ma2019hierarchical}, GRU4Rec~\citep{hidasi2016sessionbasedrecommendationsrecurrentneural}, and SASRec~\citep{kang2018self}; and (2) Generative Recommender Models, such as TIGER~\citep{rajput2023recommender}, HSTU~\citep{zhai2024actions}, and ReaRec~\citep{tang2025think}. 
Top-K Recall (R@K) and NDCG (N@K) with K=5 and 10 are used as metrics, following~\cite{rajput2023recommender}. 
Implementation
~\footnote{code and data are in \url{https://github.com/wangshy31/OneRec-Think}} 
details are in Appendix~\ref{app:implement_open}.

\subsection{Overall Performance}\label{sec:overall_performance}
The results are shown in Table~\ref{tab:MainTable}.
We could observe that models leveraging powerful reasoning-based architectures (ReaRec and our \name) consistently outperform both traditional sequential recommenders and generative recommenders.
This robust trend confirms that effective sequential prediction necessitates robust inference and contextual reasoning capabilities. 
Furthermore, \name further gets the best performance across all benchmarks. 
This significant superiority is directly attributed to explicit, text-based reasoning ability for item generation, in contrast to the more implicit, purely learned generation mechanisms in prior works.

\subsection{Ablation Study}

\begin{table}
\setlength{\abovecaptionskip}{0.05cm}
\setlength{\belowcaptionskip}{0.2cm}
\centering
\caption{Ablation Study of different variants of \name on Beauty dataset.}
\setlength{\extrarowheight}{4pt}
\label{tab:alignment_effect}
\resizebox{1.0\linewidth}{!}{
\begin{tabular}{ccccc}
\toprule
\textbf{Training Method}  & \textbf{R@5} & \textbf{R@10} & \textbf{N@5} & \textbf{N@10}  \\
\midrule
Base & 0.0460 & 0.0654 & 0.0314 & 0.0377 \\
Base+IA  & 0.0532 & 0.0735 & 0.0342 & 0.0402 \\
Base+IA+R & \textbf{0.0563} & \textbf{0.0791} & \textbf{0.0398} & \textbf{0.0471} \\
\bottomrule
\end{tabular}
}
\end{table}

We conduct an ablation study on the \textbf{Beauty} dataset, comparing three configurations: the \textit{Base} model tuned by the raw itemic token sequence, the \textit{Base+IA} model enhanced with Itemic Alignment, and the full \textit{Base+IA+R} model incorporating our enhanced reasoning mechanism, as shown in Table~\ref{tab:alignment_effect}.
It demonstrates that each component is indispensable: Itemic Alignment provides a foundational boost by creating coherent semantic representations of itemic tokens, while the reasoning mechanism yields a further significant gain, confirming that both components synergistically address core challenges in sequential recommendation.

\subsection{Industrial Experiments}  
\subsubsection{Training Settings} 
We adopt Qwen-8B~\citep{yang2025qwen3} as our backbone model, initializing its parameters from the publicly available pre-trained weights. 
The model's vocabulary is extended with 24,576 new tokens representing the three-level hierarchical itemic tokens (8,192 tokens per level), plus two special boundary tokens, \texttt{<|item\_begin|>} and \texttt{<|item\_end|>}. 
For our production environment, we implement a daily incremental training pipeline. 
The model is updated each day on a cluster of 80 flagship GPUs, processing approximately 20B tokens per day to stay current with newly generated user interaction data.
The details are shown in Appendix~\ref{app:industry_setting}.

\begin{table}[t]
    \centering
    \caption{The relative improvement of our online A/B testing on a short-video recommendation scenario.}
    \renewcommand{\arraystretch}{1.2} 
    \setlength{\tabcolsep}{5pt}      
    \begin{tabular}{l|cc}
    \toprule
     \textbf{Online Metrics} & \textbf{\name}   \\
    \hline
     App Stay Time& +0.159\%  \\
     Watch Time & +0.169\% \\
     Video View  & +0.150\% \\
     Follow  & +0.431\% \\
     Forward  & +0.758\% \\
     Like & \textcolor{gray}{+0.019\%} \\
     Collect & \textcolor{gray}{+0.098\%} \\
    \bottomrule
    \end{tabular}
    \label{online_ab_table1}
\end{table}

\subsubsection{Results}
\paragraph{Online A/B Result.}  
We deploy \name on Kuaishou, a short-video platform with hundreds of millions of daily active users.
Using a 1.29\% traffic experimental group, we compare \name with our online model for one week and report the result in Table~\ref{online_ab_table1}, where the primary metric is APP Stay Time (reflecting total user engagement time).
The primary metric, App Stay Time, shows significant gains that increase by 0.159\%.
Note that in industrial recommendation systems, 0.1\% improvements are considered substantial. 
Furthermore, interactive metrics such as Video View and Forward exhibit positive trends, indicating enhanced user engagement. 
We conducted multiple experiments at different times and consistently observed significant improvements in stay time and related interaction metrics.

\begin{table}
\centering
\caption{The Bertscore for User Understanding and Short Video Understanding Benchmark.}
\label{app:Bertscore}
\footnotesize
\resizebox{1.0\linewidth}{!}{
\begin{tabular}{c|c|c|c}
\toprule
 \multirow{2}{*}{Benchmark} & \multirow{2}{*}{Qwen3} & Qwen3 & Qwen3  \\
 & & + TW & + TW + MI \\
\midrule
User & 0.6588 & 0.6492 & \textbf{0.7053} \\
Short Video & 0.6031 & 0.6443 & \textbf{0.7300}\\
\bottomrule
\end{tabular}
}
\end{table}

\begin{figure*}[!t]
    \centering
    \includegraphics[width=1\textwidth]{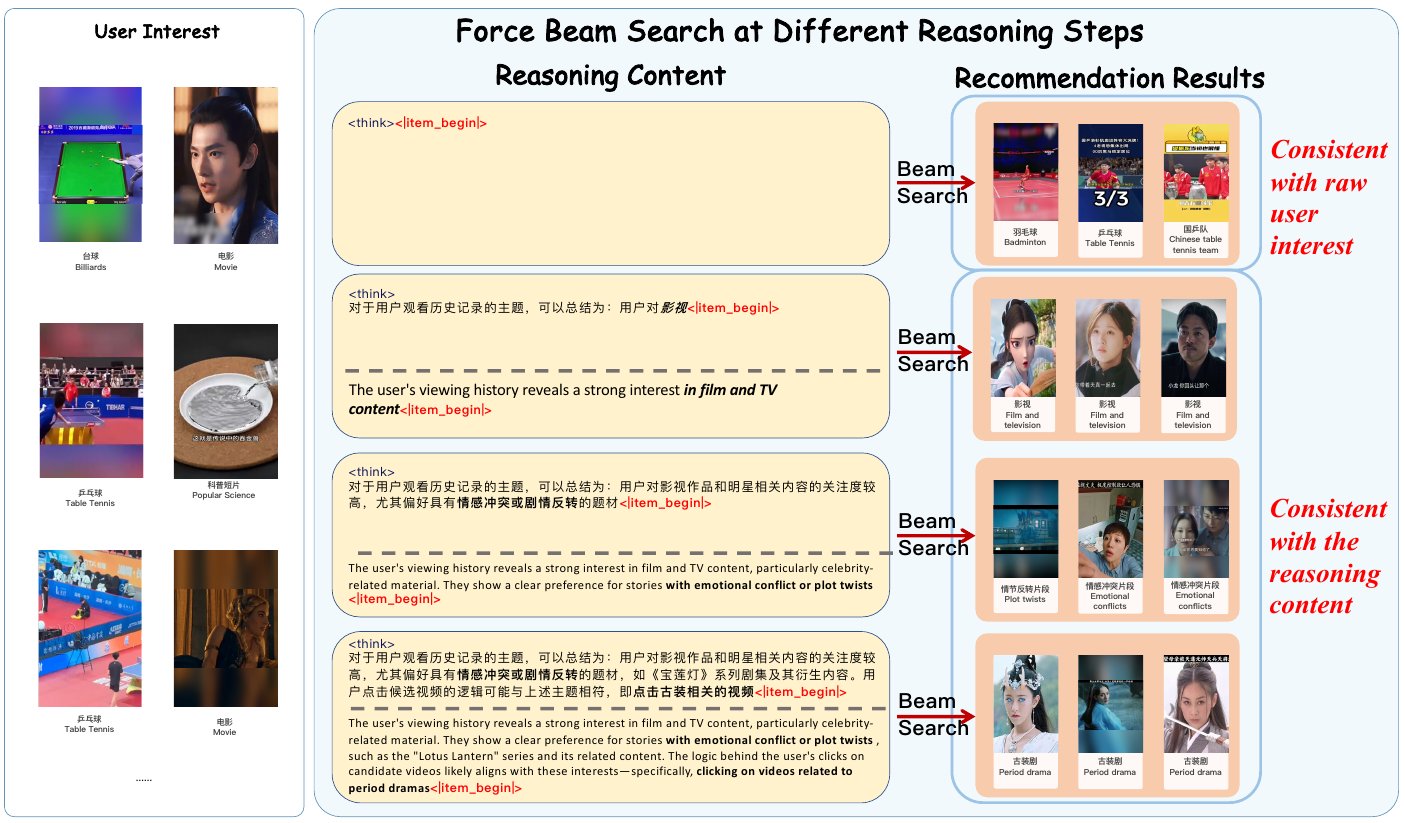}
    \caption{The model's reasoning process evolves from broad interest matching (left) to fine-grained theme specification (middle), with recommendations (right) showing semantic consistency with each reasoning step.}
    \label{fig:case_CoT_5}
\end{figure*}

\paragraph{Ablation on Itemic Alignment on industrial benchmark}
We evaluate Token Warm-up (TW) and Multi-Task Integration (MI) of the Itemic Alignment stage on our industrial User and Short Video Understanding benchmarks using BertScore~\citep{zhang2019bertscore} (details in Appendix~\ref{app:Benchmark}). 
Results in Table~\ref{app:Bertscore} reveal distinct roles for each component. 
On the text-heavy User Understanding task, TW provides limited gain over the strong Base model since the LLM can effectively process the abundant textual information directly, while MI delivers a substantial boost by translating aligned representations into actionable insights. 
In contrast, in the pure itemic token Short Video Understanding task, it shows progressive gains from both TW and MI, confirming their necessity for interpreting non-textual item information. 
These results validate that these two substages both contribute to the final performance of Itemic Alignment.

\begin{figure*}[!t]
    \centering
    \includegraphics[width=1\textwidth]{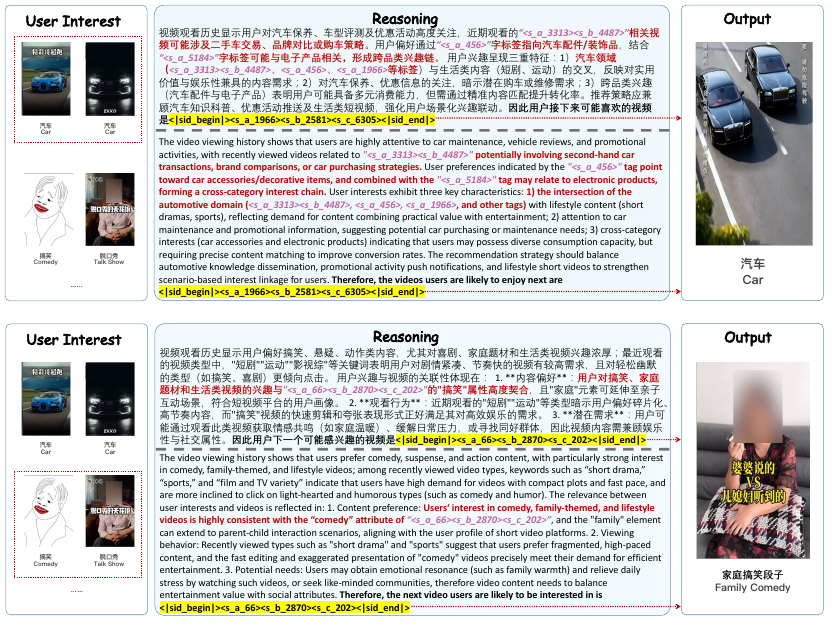}
    \caption{Demonstration of itemic-textual interleaved reasoning.}
    \label{fig:case_interleaved}
\end{figure*}

\subsection{Case Study}
Our case studies demonstrate the model's sophisticated reasoning capabilities across different scenarios. 
In conversational settings (Fig.~\ref{fig:case_CoT_3}), when the user expresses negative emotions, the model detects this affective signal and strategically shifts recommendations from general interests toward relaxing and positive content, demonstrating its ability to actively optimize the viewing experience through the interaction with the user.
In reasoning-based short-video recommendation (Fig.~\ref{fig:case_CoT_4}), the model generates diverse reasoning paths that capture fine-grained user preferences, such as specific gameplay mechanics and narrative patterns, enabling more precise recommendations beyond coarse topic matching. 
Furthermore, our consistency analysis (Fig.~\ref{fig:case_CoT_5}) reveals strong alignment between reasoning textx and recommended items when applying beam search at intermediate reasoning steps, confirming that the reasoning process genuinely guides recommendation generation rather than serving as post-hoc justification. 
Notably, our model achieves itemic-textual interleaved reasoning paths (Fig.~\ref{fig:case_interleaved}). Through precise content anchoring by itemic tokens and causal articulation by textual tokens, the interleaved reasoning delivers enhanced recommendation accuracy and transparent explanations beyond isolated modality approaches.
These results collectively validate our model's capacity for authentic, multi-faceted reasoning by demonstrating its ability to adapt to real-time interactions, capture fine-grained preferences, and maintain semantic consistency across diverse recommendation scenarios.

\section{Conclusion}
We present \name, a novel framework that bridges reasoning capabilities with generative recommendation through three key innovations: hierarchical itemic token alignment, reasoning activation via CoT supervised fine-tuning, and reinforcement-based reasoning refinement. 
Our method fundamentally transforms recommendation systems from mere item predictors into reasoning-aware models that generate interpretable rationales alongside high-quality recommendations.
Extensive experiments demonstrate that \name not only achieves state-of-the-art performance across multiple benchmarks, but also translates to concrete industrial impact with a 0.15\% gain in primary metrics like APP Stay Time.
Future work will focus on exploring user long-sequence modeling and dense RL reward for finer-grained preference modeling, further bridging LLM-based reasoning with industrial recommendation systems.

\section*{Limitations}
Despite promising empirical results, current public datasets exhibit quality constraints through their limited behavior sequence lengths and restricted item spaces. 
These limitations hinder our Reasoning Activation and Reasoning Enhancement modules from acquiring high-quality reasoning capabilities comparable to those learned from industrial-scale data. 
Consequently, we simplify and adapt our approach to achieve a stable yet simplified reasoning capacity, which remains robust within the public datasets.
To address these issues, we are actively constructing a large-scale benchmark with extended behavioral trajectories and diversified item catalogs that will enable more comprehensive evaluation of reasoning capabilities for reasoning-based recommendation models.

\section*{Ethics Statement}
In this work, we have conducted experiments for two settings: one for open-source benchmark datasets and one for the industrial scenario.
For the experiments for open-source benchmark datasets, all datasets are publicly available from previous works or public APIs while maintaining anonymity.
For the industrial scenario, we utilize user interaction data collected from our platform to train the recommendation model.
All data collection and usage strictly comply with our platform's privacy policy and terms of service, to which users have provided explicit consent.
Importantly, our training process operates solely on aggregated behavioral sequences, textual content, and user base information without accessing or processing any personally identifiable information.

\bibliography{custom}

\appendix
\section{Appendix}

\subsection{Experiment settings}\label{app:implement_open}
\paragraph{Details of Baselines}
We compare \name with competitive baselines within two groups of work, traditional recommender models and generative recommender models:
1) \textbf{BERT4Rec}~\citep{sun2019bert4rec} leverages BERT's pre-trained language representations to capture semantic user-item relationships.
2) \textbf{HGN}~\citep{ma2019hierarchical} utilizes graph neural networks to learn user and item representations for predicting user-item interactions.
3) \textbf{GRU4Rec}~\citep{hidasi2016sessionbasedrecommendationsrecurrentneural} is a lightweight graph convolutional network model focusing on high-order connections between users and items.
4) \textbf{SASRec}~\citep{kang2018self} employs self-attention mechanisms to capture long-term dependencies in user interaction history.
5) \textbf{TIGER}~\citep{rajput2023recommender} introduces codebook-based identifiers via RQ-VAE, which quantizes semantic information into code sequences for LLM-based generative recommendation.  
6) \textbf{HSTU}~\citep{zhai2024actions} reformulates recommendation problems as sequential transduction tasks
within a generative modeling framework and proposes a new architecture for streaming data.
7) \textbf{ReaRec}~\citep{tang2025think} enhances user representations through implicit multi-step reasoning within an inference-time computing framework for recommendation.
\textbf{Evaluation Metrics.} 
We use two metrics: top-$K$ Recall (R@$K$) and NDCG (N@$K$) with $K$ = 5 and 10, following~\cite{rajput2023recommender}.

\begin{figure*}[!t]
    \centering
    \includegraphics[width=.99\textwidth]{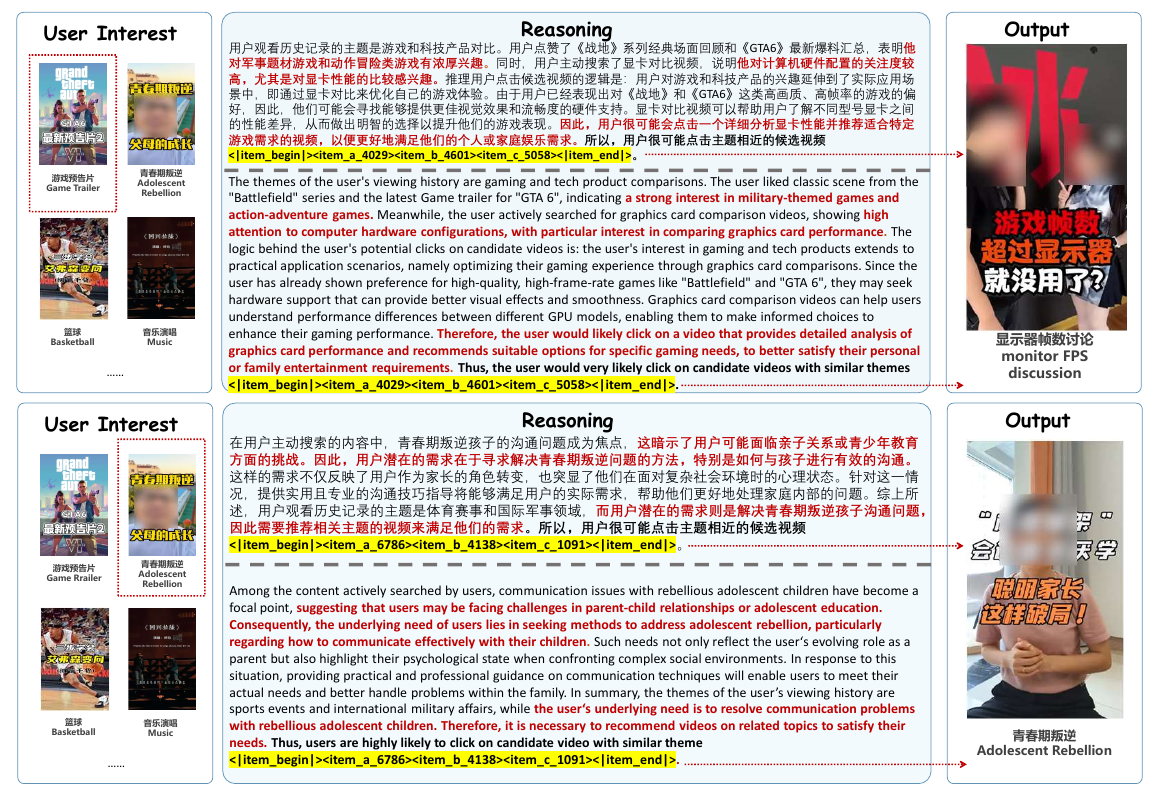}
    \caption{A reasoning example for a short-video recommendation scenario.}
    \label{fig:case_CoT}
\end{figure*}

\paragraph{Details of experiments on open-source datasets}
We adopt Qwen3-1.7B~\citep{yang2025qwen3} as our backbone model.
The model's vocabulary is extended with 1,024 new tokens representing the four-level hierarchical semantic IDs (256 tokens per level), in addition to two special boundary tokens, \texttt{<|item\_begin|>} and \texttt{<|item\_end|>}. 
All models were trained on a server equipped with flagship GPUs. 
To generate the top-K recommendations during evaluation, we employ a beam search strategy with a beam width of 10.
Given the inherent challenge of deriving a robust reasoning path $\bm{\tau}$ from the short and sparse item sequences typical of public benchmarks, we strategically employ manually constructed category-based CoT as the pruned content for Reasoning Activation to ensure stable semantic guidance.
For the training data, we adopt the pre-processing techniques from previous work~\citep{rajput2023recommender, wang2024learnableitemtokenizationgenerative}, discarding sparse users and items with interactions less than 5. 
We consider the sequential recommendation setting
and use \textit{leave-one-out} strategy~\citep{rajput2023recommender, zheng2023adapting} to split datasets. 
For training, we follow~\citep{kang2018self} to restrict the number of items in a user's history to 50.

\subsection{Model Demonstration}\label{app:cases}

\paragraph{Reasoning Cases for Short-video Recommendation}\label{app:case_CoT}

In this part, we show some cases of short video recommendations of another user.
Figure~\ref{fig:case_CoT} presents two reasoning paths generated by our model, demonstrating its capacity for multi-step interest inference and underlying need identification. 
In the first case, the model connects the user's gaming preferences with hardware comparison behaviors to deduce an unstated need for performance optimization, ultimately recommending monitor analysis videos. 
The second case reveals deeper psychological needs by associating sports/military viewing history with adolescent rebellion searches, identifying parenting challenges as the core concern. Both examples showcase our model's ability to transcend superficial topic matching and perform causal reasoning about user motivations, enabling recommendations that address both expressed interests and latent needs.

\paragraph{Semantic Comprehension Validation}
After semantic alignment, the model acquires the capability to comprehend and articulate the semantic meaning of item tokens through natural language.
To validate this emergent ability, we evaluate the capacity of the model to generate descriptive captions for itemic tokens without explicit training on this task. 
When prompted to explain what an itemic token represents, the model leverages the learned semantic correspondences to produce coherent textual descriptions that accurately capture the characteristics of the items. 
This demonstrates that the alignment process successfully establishes genuine semantic understanding rather than superficial pattern matching, as the model can now bidirectionally translate between discrete itemic tokens and their rich natural language semantics.
The following cases in Beauty datset showcase the model's caption generation capability for item semantic tokens:
\begin{simplebox}[colback=teal!5!white, colframe=teal!60!black, colbacktitle=teal!60!black]{Case: Itemic Token Explaining 1}

    \textbf{Question + Answer}
    
    \begin{lstlisting}[style=largestyle, escapechar=`, backgroundcolor=\color{teal!5!white}]
<|im_start|>system
You are a helpful assistant.<|im_end|>
<|im_start|>user
Can you explain the meaning of <|item_begin|><item_a_219><item_b_189><item_c_49><item_d_227><|item_end|>?<|im_end|>
<|im_start|>assistant
`\textbf{Its title is Manuka Honey and Manuka Oil Skin Cream, its category is Beauty > Skin Care > Face > Cleansers.}`
    \end{lstlisting}

    \textbf{Ground Truth}
    
    \begin{lstlisting}[style=largestyle, backgroundcolor=\color{teal!5!white}]
"title": "Manuka Honey and Manuka Oil Skin Cream"
"categories": "Beauty > Skin Care > Face > Creams & Moisturizers > Fluids & Lotions > Lotions"
"sid": "<|item_begin|><item_a_219><item_b_189><item_c_49><item_d_227><|item_end|>"
    \end{lstlisting}
\end{simplebox}

\begin{simplebox}[colback=teal!5!white, colframe=teal!60!black, colbacktitle=teal!60!black]{Case: Itemic Token Explaining 2}

    \textbf{Question + Answer}
    
    \begin{lstlisting}[style=largestyle, escapechar=`, backgroundcolor=\color{teal!5!white}]
<|im_start|>system
You are a helpful assistant.<|im_end|>
<|im_start|>user
Can you explain the meaning of <|item_begin|><item_a_8><item_b_18><item_c_86><item_d_131><|item_end|>?<|im_end|>
<|im_start|>assistant
`\textbf{Its title is 25 Color Professional Camouflage Makeup Palette, its category is Beauty > Makeup > Face > Concealer \& Base.}`
    \end{lstlisting}

    \textbf{Ground Truth}
    
    \begin{lstlisting}[style=largestyle, backgroundcolor=\color{teal!5!white}]
"title": "Professional 15 Color Concealer Camouflage Makeup Palette",
"categories": "Beauty > Makeup > Face > Concealers & Neutralizers",
"sid": "<|item_begin|><item_a_8><item_b_18><item_c_86><item_d_131><|item_end|>"
    \end{lstlisting}
\end{simplebox}

\subsection{Implementation Details}\label{app:industry_setting}

\subsubsection{Itemic Alignment}\label{app:SA}

\paragraph{Task details}
Here, we first introduce the aforementioned four types of tasks.

\textbf{1. Interleaved User Persona Grounding.} This part contains the interleaved itemic tokens with rich, natural-language text extracted from static user profiles. 
This process forces the model to create a robust mapping between the itemic tokens and their real-world meanings, grounded in user attributes, stated interests, and historical behaviors. 
The sample data is shown below.
\begin{simplebox}{Interleaved User Persona Grounding data}
\begin{lstlisting}[style=largestyle]
# User Profile Narrative
The user is a 25-30 year-old male based in Beijing. 

# User Profile Narrative
The user's recent searches on the platform include: "best space opera novels," and "latest NASA discoveries."

# Live Stream Behavior
He recently commented 5 times on live streams in the "Science & Tech" category.

# Like Behavior
He recently liked video <|item_begin|><item_a_1123><item_b_5813><item_c_4212><|item_end|>, captioned "Exploring the Andromeda Galaxy with the James Webb Telescope."; and video <|item_begin|><item_a_3421><item_b_8812><item_c_1234><|item_end|>, captioned "Top 10 Paradoxes of Time Travel."
...

# Comment Behavior
He recently commented "Incredible footage!" on Video: <|item_begin|><item_a_5813><item_b_1123><item_c_9876><|item_end|>, a documentary about black holes; and commented "Mind-blowing concept!" on Video: <|item_begin|><item_a_8812><item_b_3421><item_c_5432><|item_end|>, explaining the Fermi Paradox.
...

# Followed Creators
He follows creators on the platform across various fields, including science popularizers, book reviewers, and film critics specializing in sci-fi.

# User Summary
Primary Interests: This user enjoys science, especially astronomy. This user also engages with Honor of Kings content, indicating a casual gaming interest.
Secondary Interests: Diverse explorations include pet (cat) videos, traditional culture, and local food content.

\end{lstlisting}
\end{simplebox}

\textbf{2. Sequential Preference Modeling.} The model is given a sequence of a user's recent interactions (up to 256 items) and is tasked with predicting the next item. The loss is computed only on the tokens of the target itemic token.
The sample data is shown below.
\begin{simplebox}{Sequential Preference Modeling data}
\begin{lstlisting}[style=largestyle]
<|im_start|>system
You are a sequential recommendation engine. Your task is to analyze the provided sequence of user-item interactions and predict the single next item the user is most likely to engage with.
<|im_end|>
<|im_start|>user
User interaction history: <|item_begin|><item_a_1024><item_b_2048><item_c_4096><|item_end|>...<|item_begin|><item_a_1234><item_b_5678><item_c_5876><|item_end|>. Predict the next item.
<|im_end|>
<|im_start|>assistant
The next recommended item is <|item_begin|><item_a_5555><item_b_6666><item_c_7777><|item_end|>.
<|im_end|>
\end{lstlisting}
\end{simplebox}

\textbf{3. Itemic Dense Captioning.} This task asks the model to generate a textural description of a given itemic token, ensuring a robust mapping between itemic tokens and natural language. 
The loss is computed on the tokens of the target textual description.
The sample data is shown below.
\begin{simplebox}{Itemic Dense Captioning data}
\begin{lstlisting}[style=largestyle]
<|im_start|>system
You are an expert content analyst. Given an itemic token, generate a concise and accurate textual description of its content.
<|im_end|>
<|im_start|>user
Provide a description for the itemic token: <|item_begin|><item_a_1357><item_b_2468><item_c_7753><|item_end|>.
<|im_end|>
<|im_start|>assistant
This video showcases a local food stall, highlighting traditional cooking methods and signature dishes like grilled skewers to attract customers by celebrating local food culture.
<|im_end|>
\end{lstlisting}
\end{simplebox}

\textbf{4. General Language Modeling.} This part of the data contains the pretraining and instruction fine-tuning data of the general corpus to maintain the base ability of the language model.

To prepare our model for in-text reasoning recommendations, we design a two-stage training curriculum to first build a foundational semantic understanding and then develop sophisticated recommendation capabilities. 
The initial substage, \textbf{Token Warm-up}, focuses on aligning itemic tokens with their rich textual meanings.
The second substage, \textbf{Multi-Task Integration}, develops the model's core recommendation abilities while preserving the initial semantic alignment.

\paragraph{Token Warm-up}
The initial and foundational step is to imbue the model with a deep understanding of the semantic content represented by each itemic tokens ($\bm{s}_v$). 
In the open-source benchmark datsets, the dataset is consist of the user profile of the full training data.
In the industrial setting, the entire alignment process leverages a corpus of 6 billion tokens of the Interleaved User Persona Grounding task, ensuring a comprehensive semantic grounding.

During this stage, we freeze the parameters of the LLM backbone and exclusively train the embeddings of the newly introduced itemic tokens.
This focused approach allows the randomly initialized embeddings to efficiently converge to meaningful positions within the model's existing semantic space without disrupting its pre-trained knowledge. 
We use a higher learning rate of $5 \times 10^{-4}$ to train only the new token embeddings, which allows the randomly initialized embeddings to converge quickly and efficiently.

\begin{table}[!t] 
\centering
\small 
\caption{Data details for the Recommendation Enhancing.}
\label{tab:data_dist}
\begin{tabular}{ l |c }
\toprule
\textbf{Task Type} & \textbf{Data Percentage} \\
\midrule
\textbf{Interleaved User Persona Grounding} & 24.30\% \\
\midrule
\textbf{Sequential Preference Modeling} & 65.73\% \\
\midrule
\textbf{Itemic Dense Captioning} & 4.94\% \\
\midrule
\textbf{General Language Modeling} & 5.03\% \\
\bottomrule
\end{tabular}
\end{table}

\paragraph{Multi-Task Integration}\label{app:RE}
With the semantic item vocabulary successfully grounded, the second substage of our curriculum aims to develop the model's core recommendation capabilities.
Simply training on a single objective, such as next-item prediction, could cause the model to gradually treat the itemic tokens as conventional, non-semantic identifiers, thereby losing the benefits of the initial grounding substage.
To prevent this and foster a more holistic understanding, we employ the Multi-Task Integration.
The mixture of different tasks, detailed in Table~\ref{tab:data_dist}, is designed to balance the learning of collaborative patterns with the reinforcement of semantic understanding and contextual reasoning.

For the open-source benchmark dataset, we train the full parameters of the model.
For the industrial scenario, we unfreeze the model backbone and utilize LoRA~\citep{hu2022lora} for parameter-efficient fine-tuning.
We employ a more conservative learning rate of $3 \times 10^{-4}$ and fine-tune the model, which ensures a stable adaptation of the model's internal mechanisms to the recommendation tasks without disrupting its foundational knowledge.

\subsubsection{Reasoning Activation}
The Reasoning Activation stage employs a carefully orchestrated two-substage training pipeline designed to progressively develop the model's reasoning capabilities from simplified contexts to complex, noisy industrial sequences.
\paragraph{Bootstrapping with Pruned Contexts:}
In this substage, we construct a high-quality reasoning dataset by strategically pruning user behavior sequences to retain only the most relevant interactions. For each training instance, we employ a semantic similarity function $g(\cdot,\cdot)$ to extract the top-$k$ items (we set $k=10$ in our experiments) most relevant to the target interaction. 
The similarity function leverages pre-trained item embeddings to compute cosine similarity scores between historical items and the target. 
Using these pruned sequences, we prompt the semantically aligned model to generate step-by-step reasoning trajectories that explain why the target item aligns with the user's demonstrated preferences. 
The example is shown below.
\begin{simplebox}{Rationale Generation from Pruned Context}
\begin{lstlisting}[style=largestyle]
<|im_start|>system
You are an expert recommendation analyst. Given a user's relevant interaction history, provide a step-by-step reasoning process explaining why a specific target item aligns with their preferences.
<|im_end|>
<|im_start|>user
User's relevant interaction history:

<|item_begin|><item_a_1111><item_b_2222><item_c_3333><|item_end|> - Beginner's guide to sourdough bread making
<|item_begin|><item_a_2947><item_b_8156><item_c_3402><|item_end|> - Professional chef knife skills tutorial
<|item_begin|><item_a_5618><item_b_3927><item_c_7084><|item_end|> - Italian pasta recipes from scratch
<|item_begin|><item_a_7293><item_b_5471><item_c_1836><|item_end|> - Kitchen organization and meal prep tips
<|item_begin|><item_a_9105><item_b_6248><item_c_4519><|item_end|> - Comparison of stand mixers for home baking

Target item: <|item_begin|><item_a_3672><item_b_9814><item_c_2657><|item_end|> - Advanced French pastry techniques and equipment guide
Explain why this target item matches the user's interests.
<|im_end|>
<|im_start|>assistant
Reasoning: The user's viewing history shows a strong interest in advanced cooking techniques and kitchen equipment, with engagement in sourdough baking, handmade pasta, and stand mixer comparisons. The user has progressed from basic skills to time-intensive culinary projects, demonstrating commitment to mastering traditional cooking methods. French pastry represents a natural next step that combines their baking experience with technical precision. Therefore, the user would very likely click on candidate videos with similar themes <|item_begin|><item_a_3672><item_b_9814><item_c_2657><|item_end|>.
<|im_end|>
\end{lstlisting}
\end{simplebox}

\paragraph{Learning to Reason from Noisy Sequences}
Building upon the distilled reasoning patterns, this substage trains the model to transfer its reasoning capabilities to raw, unfiltered user behavior sequences. 
The training data consists of complete user histories paired with the high-quality rationales generated in Substage 1. 
We employ the joint optimization objective defined in Equation~\ref{eq:sft_loss}, which simultaneously trains the model to: (1) generate coherent reasoning trajectories by predicting rationale tokens conditioned on noisy context, and (2) produce accurate target item predictions informed by both the context and generated rationale.
The loss function applies equal weighting to both components, ensuring the model learns to perform contextual distillation while maintaining recommendation accuracy. 
The example is shown below.
\begin{simplebox}{Reasoning from Noisy Sequences}
\begin{lstlisting}[style=largestyle]
<|im_start|>system
You are an intelligent recommendation system. Given a user's complete interaction history, provide reasoning for your recommendation and then suggest the most appropriate next item.
<|im_end|>
<|im_start|>user
User's complete interaction history:
<|item_begin|><item_a_1847><item_b_6392><item_c_5104><|item_end|>, <|item_begin|><item_a_9263><item_b_4571><item_c_8026><|item_end|>, <|item_begin|><item_a_1111><item_b_2222><item_c_3333><|item_end|>, <|item_begin|><item_a_5419><item_b_8736><item_c_2948><|item_end|>, <|item_begin|><item_a_3082><item_b_7145><item_c_6379><|item_end|>, <|item_begin|><item_a_2947><item_b_8156><item_c_3402><|item_end|>, <|item_begin|><item_a_7651><item_b_2493><item_c_1087><|item_end|>, <|item_begin|><item_a_5618><item_b_3927><item_c_7084><|item_end|>, <|item_begin|><item_a_8304><item_b_9672><item_c_4215><|item_end|>, <|item_begin|><item_a_7293><item_b_5471><item_c_1836><|item_end|>, <|item_begin|><item_a_4926><item_b_1358><item_c_8740><|item_end|>, <|item_begin|><item_a_9105><item_b_6248><item_c_4519><|item_end|>, <|item_begin|><item_a_1537><item_b_8904><item_c_3261><|item_end|>
What item should be recommended next? Explain your reasoning first.
<|im_end|>
<|im_start|>assistant
<think>
The user's viewing history shows a strong interest in advanced cooking techniques and kitchen equipment, with engagement in sourdough baking, handmade pasta, and stand mixer comparisons. The user has progressed from basic skills to time-intensive culinary projects, demonstrating commitment to mastering traditional cooking methods. French pastry represents a natural next step that combines their baking experience with technical precision. Therefore, the user would very likely click on candidate videos with similar themes <|item_begin|><item_a_3672><item_b_9814><item_c_2657><|item_end|>.
</think>
Recommendation: <|item_begin|><item_a_3672><item_b_9814><item_c_2657><|item_end|>
<|im_end|>
\end{lstlisting}
\end{simplebox}

Similarly, for these tasks, we train the full parameters of the model.
For the industrial scenario, we unfreeze the model backbone and utilize LoRA for parameter-efficient fine-tuning.
We employ a learning rate of $3 \times 10^{-4}$ and fine-tune the model.

To quantify the impact of Semantic Grouding (SG) and Recommendation Enhancement (RE) in Itemic Alignment stage, we calculate the BertScore~\citep{zhang2019bertscore} between the ground truth and prediction of \name on the User Understanding and Short Video Understanding benchmarks. 
The cases of User and Short Video Understanding of Table~\ref{app:Bertscore} are shown below, and the ground truth is the accurate answer from our online multi-modal large language model.~\label{app:Benchmark}
\begin{simplebox}{Case for Short Video Understanding Benchmark}
\begin{lstlisting}[style=largestyle]
<|im_start|>system
You are an expert content analyst. Given an itemic token, generate a concise and accurate textual description of its content.
<|im_end|>
<|im_start|>user
Provide a description for the itemic token: <|item_begin|><item_a_1357><item_b_2468><item_c_7753><|item_end|>.
<|im_end|>
<|im_start|>assistant
# Answers
\end{lstlisting}
\end{simplebox}

\begin{simplebox}{Case for User Understanding Benchmark}
\begin{lstlisting}[style=largestyle]
<|im_start|>system
You are an expert content analyst. Your task is to analyze the provided user interaction history and generate a structured, actionable user profile summary..
<|im_end|>
<|im_start|>user
<|item_begin|><item_a_1111><item_b_2222><item_c_3333><|item_end|> - Beginner's guide to sourdough bread making
<|item_begin|><item_a_2947><item_b_8156><item_c_3402><|item_end|> - Professional chef knife skills tutorial
<|item_begin|><item_a_5618><item_b_3927><item_c_7084><|item_end|> - Italian pasta recipes from scratch
<|item_begin|><item_a_7293><item_b_5471><item_c_1836><|item_end|> - Kitchen organization and meal prep tips
<|item_begin|><item_a_9105><item_b_6248><item_c_4519><|item_end|> - Comparison of stand mixers for home baking
...
<|im_end|>
<|im_start|>assistant
# Answers
\end{lstlisting}
\end{simplebox}

\subsubsection{Reasoning Enhancement}
Building upon the reasoning capabilities established in the previous stage, we employ Reinforcement Learning to further refine both reasoning coherence and recommendation accuracy. 
We leverage the VERL framework~\citep{sheng2024hybridflow} for efficient optimization, which provides robust infrastructure for distributed training and scalable reward computation. 
Our training utilizes the GRPO algorithm and implements the Rollout-Beam reward within VERL. 
During training, we sample $|G|=16$ CoT paths and do beam search with width $K=32$ for each path to compute rewards, balancing exploration breadth with computational efficiency.
We train for 2 epochs with a learning rate of 1e-5, KL divergence coefficient $\beta=0.001$, clip ratio $\epsilon=0.2$ to ensure stable policy updates.

\subsubsection{System Deployment: A "Think-Ahead" Architecture}\label{app:think_ahead}
Here, we show the details of the "Think-Ahead" Architecture.
Our "Think-Ahead" architecture decouples inference into an offline stage that generates reasoning paths and initial item tokens to capture broad user intent, followed by an online stage that utilizes these tokens as constrained prefixes for real-time finalization, ensuring production-grade latency.

\paragraph{Stage~1: Reasoning-Guided Prefix Generation}

\noindent{\textit{\textbf{Reasoning Path Synthesis:}}} For user $u$ with interaction history $H_u=\mathcal{P}(\bm{s}_{v_1}, \dots, \bm{s}_{v_n})$, we sample $T$ diverse reasoning paths:
\begin{equation}
\begin{aligned}
\bm{\tau}^{(i)} &\sim P(\cdot \mid H_u; \theta),
\end{aligned}
\end{equation}
where each path $\bm{\tau}^{(i)}$  for $i \in \{1, ..., T\}$ encapsulates a distinct logical pathway connecting the user's behavioral patterns to potential interests. 

\noindent{\textit{\textbf{Constrained Prefix Generation:}}} Given our hierarchical itemic tokens with $L$ identifiers, we perform beam search to generate candidate prefixes. Here, in our industrial setting, we use $L=3$. 
For each reasoning path $\bm{\tau}^{(i)}$, we decode \textit{only the first two itemic tokens}, yielding:
\begin{equation}
\label{eq:pu}
\resizebox{0.95\columnwidth}{!}{$\displaystyle 
\mathcal{A}_u^{(i)} \!= \text{BeamSearch}\Bigl(\!P\bigl(\hat{s}_{v_{n+1}}^1,\hat{s}_{v_{n+1}}^2 \mid H_u,\bm{\tau}^{(i)};\theta\bigr),m\Bigr)
$}
\end{equation}
where $\mathcal{A}_u^{(i)}$ denotes the set of $m$ candidate item prefixes derived from path $\bm{\tau}^{(i)}$ for user u.

\noindent{\textit{\textbf{Semantic Space Materialization:}}} The union of all item prefix sets forms the user's personalized candidate space:
\begin{equation}
\mathcal{C}_u = \bigcup_{i=1}^T \mathcal{A}_u^{(i)}
\end{equation}
This set, containing $T\times m$ high-potential item prefixes, is cached in an industrial distributed storage system, effectively materializing the model's deliberative reasoning into actionable semantic hypotheses.

\paragraph{Stage~2: Prefix-Constrained Finalization}
The online stage executes rapid inference while leveraging the pre-computed semantic priors from Stage 1 upon user request arrival.

\noindent{\textit{\textbf{Constrained Decoding:}}} Upon receiving a request for user $u$, we retrieve $\mathcal{C}_u$ and infer the last token using a real-time updated OneRec following \cite{zhou2025onerec1} $h_{\text{online}}$.
The search space is restricted so that the prefix of the candidate target item $(\hat{s}_{{v_{n+1}}}^1,\hat{s}_{{v_{n+1}}}^2)$ should be in the candidate prefix set, which means  $(\hat{s}_{{v_{n+1}}}^1,\hat{s}_{{v_{n+1}}}^2)\in\mathcal{C}_u$.
Consequently, we could construct the decoding target:
\begin{equation}
\begin{aligned}
\hat{\bm{s}}_{v_{n+1}} = \underset{\bm{s}_{v_{n+1}}}{\arg\max} \ 
    & P_{h_{\mathrm{online}}}\bigl(\bm{s}_{v_{n+1}} \bigm| \bm{s}_{v_1},\dots, \bm{s}_{v_n}\bigr) \\
    & \text{s.t.} \quad (\hat{s}_{v_{n+1}}^1,\hat{s}_{v_{n+1}}^2) \in \mathcal{C}_u
\end{aligned}
\end{equation}
where the items with top-K probability are then served to the users.

In summary, our \emph{Think-Ahead} architecture presents a novel paradigm for industrial recommendation, effectively reconciling sophisticated reasoning with strict latency requirements.

\label{sec:appendix}

\end{document}